\documentclass[12pt]{article}
\usepackage{graphicx}
\usepackage{color}
\usepackage{amsmath,amssymb}
\usepackage{stmaryrd}
\begin{document}
\parindent 0pt
\baselineskip=5.5mm
\bibliographystyle{aip}
\newcommand{\be} {\begin{equation}}
\newcommand{\ee} {\end{equation}}
\newcommand{\Be} {\begin{eqnarray}}
\newcommand{\Ee} {\end{eqnarray}}
\renewcommand{\thefootnote}{\fnsymbol{footnote}}
\def\a{\alpha}
\def\b{\beta}
\def\g{\gamma}
\def\G{\Gamma}
\def\d{\delta}
\def\D{\Delta}
\def\e{\epsilon}
\def\k{\kappa}
\def\l{\lambda}
\def\L{\Lambda}
\def\t{\tau}
\def\om{\omega}
\def\Om{\Omega}
\def\s{\sigma}

\def\lg{\langle}
\def\rg{\rangle}

\def\kA{k_A}
\def\kB{k_B}
\def\kT{k_T}
\def\KA{{\bf K}_A}
\def\KB{{\bf K}_B}
\def\KX{{\bf K}_X}
\def\KY{{\bf K}_Y}
\def\WA{{\bf W}_A}
\def\WB{{\bf W}_B}
\def\WX{{\bf W}_X}
\def\WY{{\bf W}_Y}
\def\VA{{\bf V}_A}
\def\VB{{\bf V}_B}
\def\VX{{\bf V}_X}
\def\VY{{\bf V}_Y}
\def\PA{{\bf P}_A}
\def\PB{{\bf P}_B}
\def\PX{{\bf P}_X}
\def\Peq{{\bf P}^{st.}}
\def\P{{\bf P}}
\def\ProjP{{\cal P}}
\def\PiA{{\bf \Pi}_A}
\def\PiB{{\bf \Pi}_B}
\def\PiX{{\bf \Pi}_X}
\def\PiY{{\bf \Pi}_Y}
\def\GA{{\bf G}_A}
\def\GB{{\bf G}_B}
\def\GX{{\bf G}_X}
\def\GY{{\bf G}_Y}
\def\G{{\bf G}}
\def\hatG{\hat{\bf G}}
\def\W{{\bf W}}
\def\M{{\bf M}}
\def\V{{\bf V}}
\def\EinT{{\bf 1}^{\rm\!T}}

\def\koff{k_{\rm off}}
\def\kon{k_{\rm on}}
\def\fext{{\rm f}_{\rm ext}}
\def\Keff{K_{\rm eff}}

\def\KmA{K_{\rm m}^{(A)}}
\def\KmB{K_{\rm m}^{(B)}}
\def\KmX{K_{\rm m}^{(X)}}
\def\KmT{K_{\rm m}^{(T)}}
\newcommand{\tred}[1]{\textcolor{red}{#1}}

\noindent
\begin{center}
{\Large {\bf Statistics of reversible transitions in two-state trajectories in force-ramp spectroscopy} }\\
\vspace{0.5cm}
\noindent
{\bf Gregor Diezemann} \\
{\it
Institut f\"ur Physikalische Chemie, Johannes Gutenberg Universit\"at Mainz,\\
Duesbergweg 10-14, 55128 Mainz, FRG
\\}
Revised Version
\end{center}
\vspace{1cm}
\noindent
{\it
A possible way to extract information about the reversible dissociation of a molecular adhesion bond from force fluctuations observed in force ramp experiments is discussed.
For small loading rates the system undergoes a limited number of unbinding and rebinding transitions observable in the so-called force versus extension (FE) curves.
The statistics of these transient fluctuations can be utilized to estimate the parameters for the rebinding rate.
This is relevant in the experimentally important situation where the direct observation of the reversed FE-curves is hampered, e.g. due to the presence of soft linkers.
I generalize the stochastic theory of the kinetics in two-state models to the case of time-dependent kinetic rates and compute the relevant distributions of characteristic forces.
While for irreversible systems there is an intrinsic relation between the rupture force distribution and the population of the free-energy well of the bound state, the situation is slightly more complex if reversible systems are considered.
For a two-state model, a 'stationary' rupture force distribution that is proportional to the population can be defined and allows to consistently discuss quantities averaged over the transient fluctuations.
While irreversible systems are best analyzed in the soft spring limit of small pulling device stiffness and large loading rates, here I argue to use the stiffness of the pulling device as a control parameter in addition to the loading rate.
}

\vspace{0.5cm}
\noindent
PACS numbers: 82.37.Np, 82.37.Rs, 87.10.Mn, 87.15.Fh
\vspace{1cm}
\section*{I. Introduction}
The strength of non-covalent adhesive bonds of molecular complexes like biomolecules or adhesion clusters can routinely be determined employing various techniques of force spectroscopy, see e.g.\cite{Evans:1997, Evans:1998, Evans:2001, Merkel:2001, Ritort:2006}.
Usually, one end of the molecular complex is fixed spatially and another part is pulled away from it with a constant pulling velocity along a prescribed direction, i.e. applying a force ramp.
The analysis of the observed distribution of rupture forces allows to gain valuable information about the free energy landscape of the binding pocket and the kinetics of bond rupture\cite{Bell:1978, Dudko:2006, Dudko:2008}.

Very often rebinding only plays a minor role and bond rupture appears to be an irreversible process that can be understood in terms of the model developed by Bell\cite{Bell:1978} in the simplest case.
The assumption is that the energy barrier to bond rupture is reduced via the application of an external force and the rate of unbinding is given by Kramers theory\cite{Hummer:2003}.
In many theoretical treatments it is assumed that the stiffness of the pulling device is small compared to the curvature of the free energy landscape of the bound state.
In this limit, only the loading rate $\mu$, the product of the pulling velocity $V$ and the effective force constant of the pulling device, 
$\Keff$, is relevant for the analysis of the rupture events.
Beyond this soft spring limit one has to modify the models and the stiffness $\Keff$ becomes an important separate parameter\cite{G67, Tshiprut:2008, Maitra:2010}.

Besides irreversible bond rupture also the behavior of systems exhibiting reversible rebinding has been investigated by force spectroscopy.
Important experimental examples showing reversible behavior are provided by certain 
biomolecules\cite{Liphardt:2001, Manosas:2006, Bornschlogl:2006} and also by specially designed molecular complexes\cite{G69}.
Also theoretically, reversible rebinding has been investigated for the case of force ramp spectroscopy (FRS)\cite{G67, 
Seifert:2002, Li:2006,Zhang:2013}.
Here, the rupture force distributions have been related to the populations of the wells of the assumed double well potential in a phenomenological way and the statistics of the transitions among the two states has not been considered in detail.
Apart from these considerations starting from double well potentials very detailed modeling of the two-state kinetics of biomolecules has been presented\cite{Ritort:2002, Manosas:2005}.
In particular, the refolding transitions that can be observed as fluctuations in the FE-curves has been considered for the first time and the number of these transitions has also been the subject of experimental investigations\cite{Manosas:2009}.
Furthermore, the importance of reversible rebinding in near equilibrium situations has been 
demonstrated\cite{Friddle:2008,Friddle:2012,Noy:2013}.

Based on the general theory of single molecule two-state trajectories that has been developed over the last 
decade\cite{Cao:2000,GS:2006,Flomenbom:2008} we have treated the statistics of reversible rebinding in the context of force-clamp spectroscopy (FCS)\cite{G68, G72} where a constant force is applied to the system.
In particular, the differences of various counting schemes have been discussed and it has been pointed out that it should be possible to monitor deviations from simple Markovian dynamics due to the fact that via the application of a constant external force a wide range of equilibrium constants is accessible.

In the present paper, I will treat the statistics of two-state trajectories for kinetic rates that explicitly depend on time, as is the case if one treats FRS.
After reviewing the general theory and the introduction of the relevant distribution functions I will consider the simple case of a phenomenological two-state system.
For this example, the quantities that are relevant in the analysis of force versus extention (FE) curves monitored in FRS are computed.
If it is possible experimentally to observe the rebinding explicitly by inverting the pulling direction after the rupture event, one can directly extract the kinetic rates for rebinding and information about the free energy minimum corresponding to the open/unbound state from the distribution of rejoin forces\cite{G67}.
Due to the small achievable loading rates laser optical tweezers appear well suited to observe reversible rebinding in particular when nucleic acids are considered.
On the other hand the use of atomic force microscopes allows the determination of larger forces\cite{Zoldak:2013}.
This is particularly interesting when supramolecular structures like specially designed 
molecules\cite{G69,Lussis:2011} or also foldamers\cite{Liu:2013} are investigated.
However, often the impact of polymeric linkers prevents the observation of rebinding in the relax mode and one mainly observes rebinding transitions in the FE-curves monitored in the pull mode as strong fluctuations in the force.
I will discuss the possibility to obtain relevant kinetic informations from these fluctuations via variation of the stiffness of the pulling device.
\section*{II. Theory}
In this section I will briefly recall the theory of the statistics of two-state trajectories as observed in single molecule spectroscopy.
Special emphasis will be given to time-dependent transition rates among the conformational states under consideration.
In particular, the relation of the transition-time distributions (or rupture force distributions) and the probability of finding the system in a given state with a prescribed history will be analyzed in detail.
While the quantities that are most relevant to FCS have been discussed in detail in ref.\cite{G72}, in the present paper I will concentrate on FRS where a constant pulling velocity is used to separate different parts of a molecular complex.
In this case, the transition rates explicitly depend on the measuring time in terms of the applied force:
\be\label{f.mut}
f=\mu\times t
\quad\mbox{with}\quad\mu=V\times\Keff
\ee
where $\mu$ denotes the loading rate.
Note that $\Keff$ is meant to consist of the bare spring constant of the pulling apparatus and the compliance of the linker used to manipulate the molecular complex under consideration.
In a simple model, $\Keff$ is represented by a connection in series of the two relevant 
springs\cite{G67,Tshiprut:2008,Noy:2013}.
\subsection*{A. Two-state statistics for time-dependent rates}
As in ref.\cite{G72}, the discussion will concentrate on a system that is defined by two sub-ensembles $A$ and $B$, each of which possibly consists of a multitude of (conformational) substates.
The populations of these ensembles are described by the vectors $\PX^{\rm T}=(p_{X,1},p_{X,2},\cdots,p_{X,N_X})$ where the dimensions $N_X$ give the number of substates in the X-ensemble, $X=A$, $B$.
Assuming a Markovian dynamics, the corresponding master equation\cite{vkamp} reads:
\Be\label{ME.AB}
{d\over dt}
\left(\begin{array}{c}\PA(t)\\
\PB(t)\end{array} \right)
=\left(\begin{array}{cc}
\WA(t) & \VB(t)\\
\VA(t) & \WB(t)	
\end{array} \right)
\left(\begin{array}{c}\PA(t)\\
\PB(t)\end{array} \right)
\Ee
with $\WX(t)=\PiX(t)-\KX(t)$.
The elements of the matrices $\VX(t)$ are the inter-ensemble transition rates $k_{Y,\a;X,\a'}(t)$ for each $(X,\a')\to(Y,\a)$ transition ($\a=1,\cdots,N_X$), $\left(\VX(t)\right)_{\a,\a'}=k_{Y,\a;X,\a'}(t)\quad (X\neq Y)$
and $\KX(t)$ consists of the diagonal part of these matrices, 
$(\KX)_{\a,\a'}(t)=\d_{\a,\a'}\sum_{\a''}k_{Y,\a'';X,\a}(t)$.
The so-called exchange matrices $\PiX(t)$ consist of the rates $\g_{X;\a,\a'}(t)$ for the transitions 
$(X,\a')\to(X,\a)$ within the $X$-ensembles,
$\left(\PiX(t)\right)_{\a,\a'}=-\d_{\a,\a'}\sum_{\a''\neq\a}\g_{X;\a'',\a}(t)+(1-\d_{\a,\a'})\g_{X;\a,\a'}(t)$.
The matrices just introduced obey the following sum-rules, which hold for arbitrary time-dependences of the rates\cite{vkamp}:
\be\label{VW.Props}
\EinT\W(t)=0
\quad;\quad
\EinT_X\PiX(t)=0
\quad;\quad
\EinT_Y\VX(t){\bf A}_X=-\EinT_X\WX(t){\bf A}_X
\ee
where ${\bf A}_X$ denotes an arbitrary vector of dimension $N_X$. 
Furthermore, $\EinT=(1,\cdots,1)$ is a summation row vector of dimension $(N_A+N_B)$ and $\EinT_X$ is the corresponding vector restricted to the dimension $N_X$.
These relations are very useful in a number of the algebraic manipulations of the expressions of the various quantities discussed in the present paper.
\\
The corresponding Green's function obeys the same master equation as the populations:
\be\label{G.ME}
\dot\G(t,t_0)=\W(t)\G(t,t_0)
\ee
where $\W(t)$ is the matrix defined in eq.(\ref{ME.AB}) and $\dot\G(t,t_0)$ denotes the derivative with respect to the later time $t$, $\dot\G(t,t_0)=\partial\G(t,t_0)/\partial t$.
The initial condition is given by $\G(t,t)={\bf E}$, where ${\bf E}$ denotes the unit matrix.

For a treatment of the statistics of the $A\leftrightarrow B$-transitions, one decomposes the transition-rate matrix according to\cite{GS:2006}:
\be\label{W.decomp}
\W(t)=\W_0(t)+\V(t)
\ee
Here, $\W_0(t)$ consists of the rates for unobserved transitions, in our case the intra-ensemble transitions, and the elements of $\V(t)$ are the rates for the observed transitions:
\be\label{VW.events}
\W_0(t)=\left(\begin{array}{cc}\WA(t) & 0\\
0 & \WB(t)	\end{array} \right)
\quad;\quad
\V(t)=\left(\begin{array}{cc}0 & \VB(t)\\
\VA(t) & 0	\end{array} \right)
\ee
This decomposition has been termed 'event-counting' in ref.\cite{G72} and one has for the 'unperturbed' Green's functions:
\be\label{Gprime.ev}
\G_0(t,t_0)=\left(\begin{array}{cc}\GA(t,t_0) & 0\\
0 & \GB(t,t_0)	\end{array} \right)
\quad\mbox{with}\quad
\dot{\bf G}_X(t,t_0)=\WX(t)\GX(t,t_0)
\ee
Several statistical properties of the system described this way for time-independent transition rates have been discussed in detail in ref.\cite{G72} and in the literature dealing with the statistics of two-state trajectories more generally, with particular focus on the analysis of the waiting time distributions\cite{Cao:2000,FKS:2005,FS:2008}.

In the present paper, I will concentrate on the situation that often is encountered in FRS on reversibly bonded systems.
In the most simple situation, one observes a transition from the closed ($A$) conformation to the open ($B$) state at a force at which the rates for the $A\to B$- and the $B\to A$-transitions are very similar. Further increasing the force leads to a situation in which the system will eventually stay in the $B$-ensemble and not return to $A$.
However, for intermediate forces it is very likely that one observes a number of 'back and forth' transitions\cite{Evans:1997}.
The statistical properties of these transitions can be described by the probability to observe $N$ transitions in the time $t$, starting from an equilibrium population in the $A$-ensemble.
In the following description I will mainly follow the treatment of ref.\cite{GS:2006} with the changes needed to adapt the theory to the present situation of time-dependent rates.

The probability to find the system in $X$ after $N$ transitions starting from an arbitrary population $\P(t_0)$ is given by:
\be\label{PXN.def}
P_X(N|t)=\ProjP_X \left\{\G_0(t,t_N)\left[*\V(t_N)\G_0(t_N,t_0)\right]^N\P(t_0)\right\}
=\ProjP_X\left\{\P_X(N|t,t_0)\P(t_0)\right\}
\ee
Here, $\ProjP_X$ projects onto the $X$-ensemble, i.e. $\ProjP_A=({\bf 1}_A,{\bf 0}_B)^{\rm\!T}$ and 
$\ProjP_B=({\bf 0}_A,{\bf 1}_B)^{\rm\!T}$.
Furthermore, $\G_0(t,t_N)\left[*\V(t_N)\G_0(t_N,t_0)\right]^N$ is a $N$-fold convolution with the outermost internal time variable denoted by $t_N$, for instance
\Be
\P_X(2|t,t_0)
=\int_{t_0}^t\!dt_2\int_{t_0}^{t_2}\!dt_1\G_0(t,t_2)\V(t_2)\G_0(t_2,t_1)\V(t_1)\G_0(t_1,t_0)
\nonumber
\Ee
Note, that in the definition of $P_X(N|t)$ the dependence on the initial time $t_0$ is to be understood implicitly.
In the present paper, usually $t_0=0$ will be used.
Knowledge of the $P_X(N|t)$ completely determines the statistics of the transitions in the system.

The relation of the $P_X(N|t)$ to the populations $p_X(t)=\ProjP_X\left\{\G(t,t_0)\P(t_0)\right\}=\EinT_X\PX(t)$, the solution of the master equation (\ref{ME.AB}), is given by:
\be\label{pX.Sum.PXN}
p_X(t)=\sum_{N=0}^\infty P_X(N|t)
\ee
as is obvious intuitively and can be derived using the generating functional discussed in Appendix A.
The relevance of this expression becomes evident when considering FE curves for small loading rates, where the system is near equilibrium. 
As we have pointed out earlier, in this situation the data are most easily analyzed in terms of averged FE curves\cite{G67}.
Then, the 'stationary' distribution of rupture forces is relevant.
The question remains how to relate the populations $p_X(t)$ to such distributions.
Generally, the corresponding (rupture) force distributions of course also depend on the history of the transitions.
In particular, the probability distribution of the times $\t$ until, starting from an arbitrary initial state, a 
$X\to Y$-transition takes place after $n$ intermediate transitions is given by:
\be\label{rhoX.def}
\rho_X(\t|n)=\ProjP_Y\left\{\V(\t)\P_X(n|\t,t_0)\P(t_0)\right\}
\ee
where the matrix $\P_X(n|t,t_0)$ is defined in eq.(\ref{PXN.def}).
If time is transformed into force, as in eq.(\ref{f.mut}), the resulting quantities $\rho_X(f|n)$ are the respective (rupture) force distributions.
These distributions of course coincide with properly defined (non-equilibrium) waiting time distributions, for instance
\[
\rho_X(t|0)=\Psi_X(t)=\EinT_Y\left\{\VX(t)\G_X(t,t_0)\P_X(t_0)\right\}
\]
which is the distribution of life-times in the $X$-ensemble.
The distribution of 'turn-over 'times\cite{G72} is given by 
\[
\rho_X(t|1)=\int_{t_0}^{t}\!dt_1\Psi_{Y,X}(t_1,t)
=\int_{t_0}^{t}\!dt_1\EinT_Y\left\{\VY(t)\G_Y(t,t_1)\VX(t_1)\G_X(t_1,t_0)\P_X(t_0)\right\}
\]

Generally, relations between the quantities $P_X(n|t)$ and $\rho_X(t|n)$ allow to define the distribution functions relevant for the interpretation of FRS data.
However, even though both quantities depend on the matrix $\P_X(n|t,t_0)$, the derivation of such relations is not trivial and 
can be obtained using the respective generating functional.
The actual calculation is presented in Appendix A and here I only quote the result:
\be\label{PNX.aus.RhoNX}
\dot P_X(N|t)=-\rho_X(t|N)+\rho_Y(t|N-1)
\ee
and
\Be\label{RhoNX.aus.PNX}
\rho_X(t|2n)
&&\hspace{-0.6cm}=
-\sum_{k=0}^n\left[\dot P_X(2k|t)+\dot P_Y(2k-1|t)\right]
\nonumber\\
\rho_X(t|2n+1)
&&\hspace{-0.6cm}=
-\sum_{k=0}^n\left[\dot P_X(2k+1|t)+\dot P_Y(2k|t)\right]
\Ee
where it is to be understood that $\rho_X(t|n)=0$ and $P_X(n|t)=0$ for $n<0$.
One thus has
\be\label{Rho0X.aus.P0X}
\rho_X(t|0)=-\dot P_X(0|t)
\ee
as can be seen from the definitions, eqns.(\ref{PXN.def}) and (\ref{rhoX.def}).
Therefore, a simple relation between a force distribution and a probability to find the system in one of the states only exists for the first transition.
In the general case, it does not seem to be possible to simplify the above relations further and the existence of dynamic disorder or intermediate states complicates the analysis. 
Additionally, the definition of some 'stationary' transition time distribution or rupture force distribtuion relating to the populations $p_X(t)$ is not straightforward.
Eqns.(\ref{PNX.aus.RhoNX}) and (\ref{RhoNX.aus.PNX}) are the most important results of the present analysis.
In the following, I will simplify the system such that analytical results emerge that can be used in the interpretation of FRS experiments.
\subsection*{B. The phenomenological two-state model}
If the time-scale for the intra-ensemble transitions ($\g_{X;\a,\a'}(t)$) is much faster than that for the inter-ensemble transitions ($k_{Y,\a';X,\a}(t)$), one can map the above model on a pure two-state model (TSM), in which each ensemble consists of a single state.
This means that in the general expressions given above, one only has a single rate $k_{Y;X}(t)$ instead of the various 
$k_{Y,\a;X,\a'}(t)$ and accordingly also $k_X(t)=k_{Y;X}(t)$ instead of $k_{X;\a}(t)=\sum_{\a'}k_{Y,\a';X,\a}(t)$.
The matrices consisting of the various transition rates reduce to $\VX(t)\to k_X(t)$ and $\WX(t)\to[-k_X(t)]$.
In this case, the master equation, eq.(\ref{ME.AB}), simplifies to
\be\label{ME.TSM}
\dot p_X(t)=-k_X(t)p_X(t)+k_Y(t)p_Y(t)\quad;\quad Y\neq X
\ee
The probabilities $P_X(N|t)$ can be calculated according to eq.(\ref{PXN.def}) using the above relations along with 
$\ProjP_A=(1,0)^{\rm\!T}$ and $\ProjP_B=(0,1)^{\rm\!T}$ (remember that there is only a single state in each ensemble).
Alternatively, one can use the generating functional $F_X(z,t)$ in order to show the following simple relation:
\be\label{PXNt.TSM}
\dot P_X(N|t)=-k_X(t)P_X(N|t)+k_Y(t)P_Y(N-1|t)
\ee
This equation can be understood intuitively, as the first term on the right hand side gives the loss of probability and the second stems from transitions from the other state, in which the system resides prior to the actual transition.
If one now assumes for the moment that the rates are time-independent,
$k_X(t)=k_X$, one finds the analytical expressions for the $P_X(N|t)$ that have been given earlier in the literature on two-state systems\cite{BW:2002,GS:2003,JP:2009}.
In ref.\cite{G68} we have considered $P_X(N|t)$ for simple models of dynamic disorder.

If the rates are time-dependent as in the case of force ramp spectroscopy, one has for example:
\be\label{PA0.PB1}
P_A(0|t)=G_A(t,t_0)
\quad\mbox{and}\quad
P_B(1|t)=\int_{t_0}^tdt_1G_B(t,t_1)k_A(t_1)G_A(t_1,t_0)
\ee
where the Greens functions according to eq.(\ref{Gprime.ev}) is given by
$G_X(t_2,t_1)=\exp{\![-\int_{t_1}^{t_2}\!dtk_X(t)]}$ and it has been assumed that the system occupied state $A$ in the beginning, $p_A(0)=1$ and $p_B(0)=0$.
The expressions for higher order probabilities follow in the same way from eq.(\ref{PXNt.TSM}).

In the phenomenological two-state model, in contrast to the more general case treated in the last section, one can give a simple explicit expression relating the distributions $P_X(n|t)$ and 
$\rho_X(t|n)$:
\be\label{RhoXN.PXN.TSM}
\rho_X(t|n)=k_X(t)P_X(n|t)
\ee
This important relation is obtained by either using the definition of $\rho_X(t|n)$, eq.(\ref{rhoX.def}), with the above mentioned substitutions for the various matrices for the phenomenological two-state model.
Another way of obtaining eq.(\ref{RhoXN.PXN.TSM}) is to insert eq.(\ref{PXNt.TSM}) into 
eq.(\ref{RhoNX.aus.PNX}).
On the other hand, one can also take an intuitive way to derive eq.(\ref{RhoXN.PXN.TSM}). 
Recognizing that all events are independent of each other in the TSM, one has two factors defining $\rho_X(t|n)$.
First, the probability to find the system in $X$ at time $t$ after exactly $n$ transitions is just $P_X(n|t)$.
The probability for a ($X\to Y$)-transition between $t$ and $(t+dt)$ is given by $k_X(t)dt$ and thus one finds 
$\rho_X(t|n)dt=k_X(t)P_X(n|t)dt$, i.e. eq.(\ref{RhoXN.PXN.TSM}). 

Eq.(\ref{RhoXN.PXN.TSM}) gives a simple relation between the distribution of transition times and the 'populations' of the respective state. 
As mentioned above, the existence of such a relation is not a trivial consequence of Markovian dynamics, but additionally the statistics of the transitions has to be distributed in Poisson-like manner.
As will be discussed below, eq.(\ref{RhoXN.PXN.TSM}) along with eq.(\ref{pX.Sum.PXN}) allows the definition of a stationary transition time distribution.
The relation to the well known expressions used in the analysis of FRS-data of irreversible systems\cite{Dudko:2006,Hummer:2003} is obtained recognizing that in this case one has $k_Y(t)=0$.
Then, from eqns.(\ref{ME.TSM}) and (\ref{PXNt.TSM}) it is evident that $\dot P_X(0|t)=\dot p_X(t)$ and thus
the life-time distribution of the state $X$ coincides with the population.
\section*{III. Force ramp spectroscopy}
The general results obtained above will be applied to the situation typically encountered in FRS.
As mentioned already in the Introduction, I will solely consider the so-called pull mode, meaning that the system is stretched by the external mechanical force and the FE-curves are monitored. 
The possibility of observing rebinding via inversion of the pulling direction in the 'relax mode' appears to be limited to special systems. 
Therefore, it is interesting to examine under which conditions it is possible to extract informations about the open $B$ state from  FE-curves conducted in the pull mode.

In the case of a linear dependence of the applied force on time as given in eq.(\ref{f.mut}), $f=\mu\times t$, one can easily convert from time to force in the description of the system.
As mentioned above, in the present paper, the impact of flexible linkers will be condensed in an effective force constant of the pulling device, $\Keff$.
Additionally, I will concentrate on the simple phenomenological two-state model in all calculations.

Given a particular force-dependence of the transition rates, $k_X(f)$, one can calculate the $G_X(f,f_0)$ and from these the quantities
$P_X(N|f)$ according to eq.(\ref{PXNt.TSM}) after transforming time to force.
The distribution of transition times is transferred into corresponding force distributions and according to eq.(\ref{RhoXN.PXN.TSM}) is simply given by 
\[
\rho_X(f|N)=\mu^{-1}k_X(f)P_X(N|f)
\]
In the following, I will present some illustrative calculations and discuss how the present approach can possibly be utilized to extract information about the rebinding process from FE-curves.
In all calculations, the phenomenological Bell model\cite{Bell:1978} is assumed to be applicable to model the force dependence of the transition rates, one has
\be\label{KX.Bell}
k_A(f)=k_A(0)e^{\a_Af}
\quad\mbox{and}\quad
k_B(f)=k_B(0)e^{-\a_Bf}
\ee
where $(\a_X/\b)$ ($\b=1/T$ with the Boltzmann constant set to unity) denotes the effective distance of the respective potential well from the transition state.
The Greens functions then are simply given by $G_A(f,f_0)=e^{-[k_A(f)-k_A(f_0)]/(\a_A\mu)}$ and 
$G_B(f,f_0)=e^{[k_B(f)-k_B(f_0)]/(\a_B\mu)}$.
The expressions for the kinetic rates in eq.(\ref{KX.Bell}) are considered here as a limiting case of the rates obtained in a Kramers like fashion assuming local harmonic approximations for the minima and the transition state\cite{G67}, cf. Appendix B.
In the present paper, I will assume large force constants $\KmT\gg\Keff$ and $\KmA\gg\Keff$
so that $\a_A\simeq\b(q_T-q_A)$ is the bare distance of minimum $A$ (located at $q_A$) from the transition state (at $q_T$), cf. eq.(\ref{alfa.LH}).

It is meaningful to assume that the free energy minimum of the 'open' structure (located at $q_B$) is less stiff and therefore the dependence of $\a_B$ on the stiffness of the pulling device, $\Keff$, will be kept, cf. eq.(\ref{XiX.def}):
\be\label{alfa.B}
\a_B=\b(\xi_Bq_B-q_T)
\quad\mbox{with}\quad\xi_B=\left[1+\Keff/\KmB\right]^{-1}
\ee
We have discussed this dependence in ref.\cite{G67} and it has also been considered in a model to account for deviations from the soft spring limit\cite{Maitra:2010}.
Of course, more sophisticated models can be employed but for the present purpose of discussing the qualitative features of the distributions this choice is sufficient. 
In particular, it will be demonstrated that similar to the situation in irreversible bond 
rupture\cite{Maitra:2010} it might be helpful in the analysis of reversible systems to vary the stiffness of the pulling device in order to extract valuable informations about the energy landscape of the system, such as 
$\KmB$.

\begin{figure}[h!]
\center
\includegraphics[width=7.0cm]{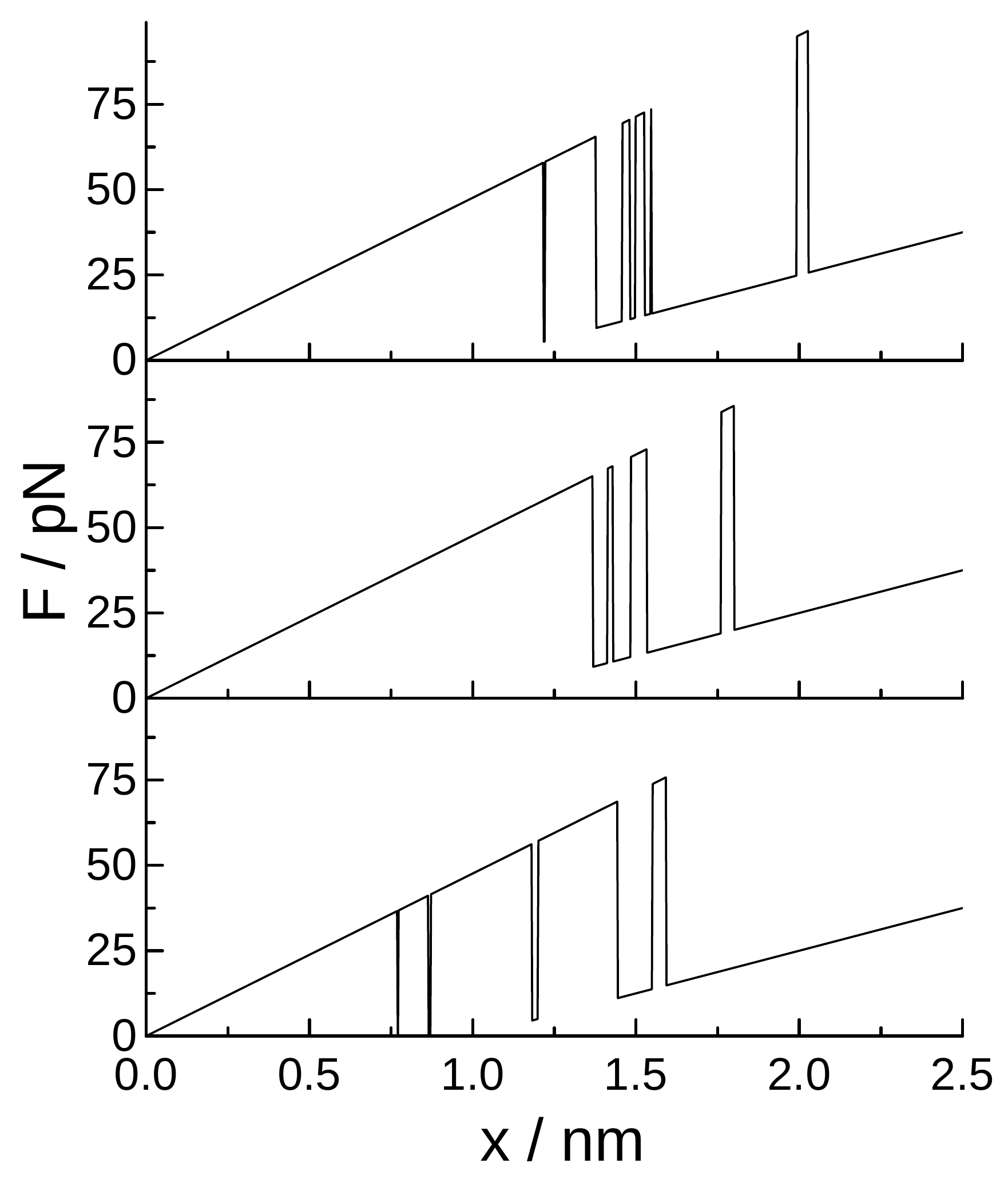}
\includegraphics[width=7.0cm]{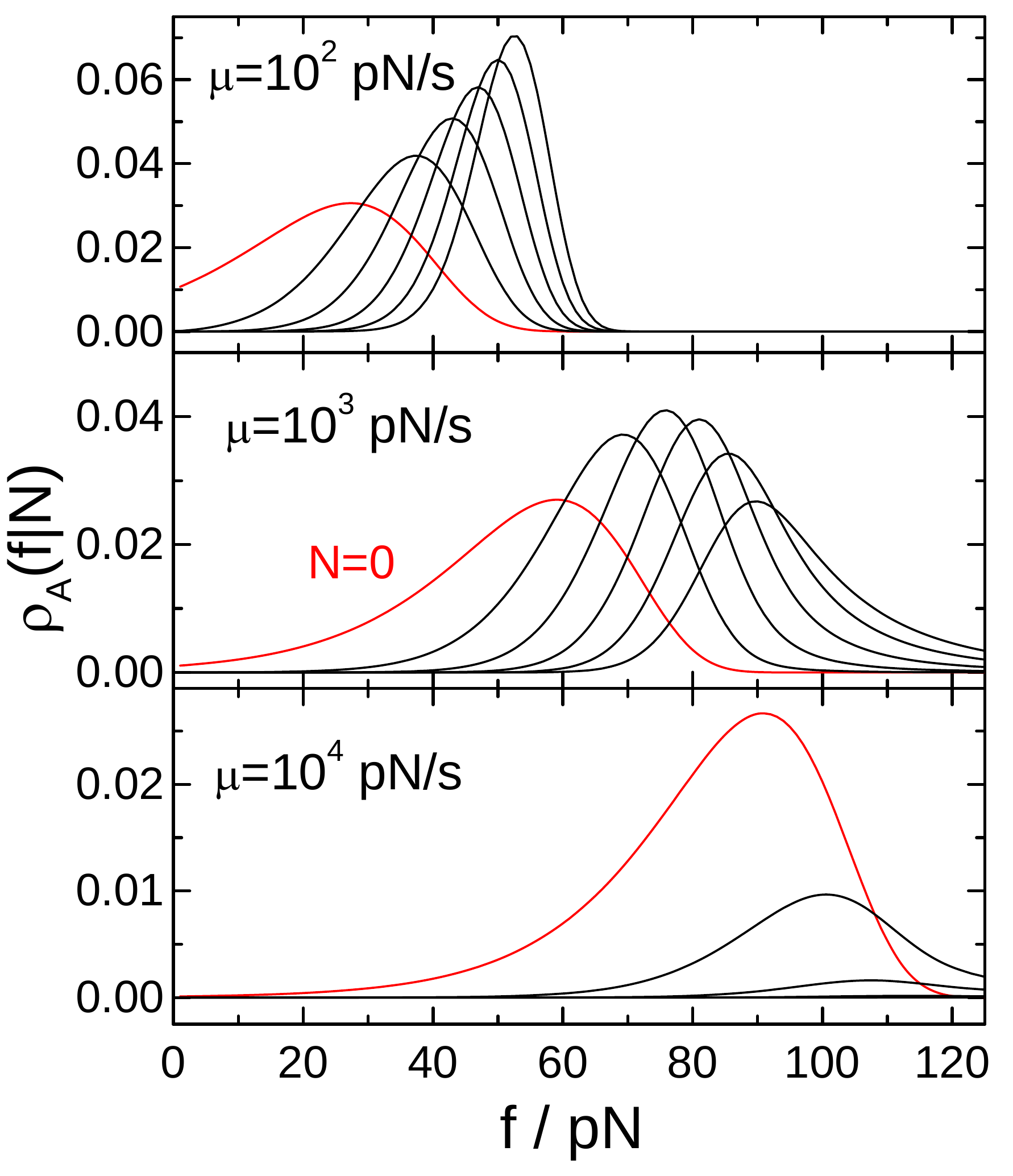}
\vspace{-0.25cm}
\caption{a) (left panel) Representative FE-curves ($x=V\cdot t$) for $\Keff=50$ pN/nm and $\mu=10^3$ pN/s.
b) (right panel) $\rho_A(N|f)$ versus force ($f=\mu\cdot t$) for $\Keff=50$ pN/nm and $N=0,\,2,\,4,\,6,\,8,\, 10$ from left to right.
The remaining parameters are: The positions of the extrema are: $q_A=0$, $q_T=0.3$ nm and $q_B=1.0$ nm.
The curvatures are $\KmA=10^3$ pN/nm and $\KmB=50$ pN/nm ($\KmT$ is assumed to large).
The intrinsic rates are: $k_A(0)=1$ s$^{-1}$ and $k_B(0)=10^4$ s$^{-1}$.}
\label{Fig.one}
\end{figure}

In Fig.\ref{Fig.one}a, representative FE-curves obtained from kinetic Monte Carlo (KMC) simulations of the 
TSM\cite{Rief:1998} are shown, where in the beginning the system resides in well $A$.
The slopes used to model the FE-curves are those resulting from the simple Gaussian approximation discussed in ref.\cite{G67}, cf. Appendix B, eq.(\ref{dyn.strength}).
The slope of the line 'in well $X$' is given by $\xi_X\Keff$ and the intercept with the force axis is at 
$(-\xi_X\Keff\cdot q_X)$.
This approximation represents the results of Brownian dynamics simulations to an excellent approximation, cf. 
ref.\cite{G67}.
If soft linkers are included, the FE-curves will be linear only in the near vicinity of the transitions and this has to be included in the analysis\cite{Hummer:2003}.
For a given loading rate, the KMC simulations show that the number of observable unbinding and rebinding transitions is a strongly fluctuating quantity.
The largest number of back and forth transitions is observed in the force regime where the equilibrium constant $k_B(f)/k_A(f)$ is on the order of unity.
For the parameters used this is for $f\sim 75$ pN.

Fig.\ref{Fig.one}b shows $\rho_A(f|N)$ for various $N$ and different loading rates $\mu$. 
Due to the strong fluctuations in the individual curves, a large number of FE-curves has to be analyzed in order to extract the distributions.
It is apparent that with increasing loading rate less back and forth transitions are observed and the maxima of the distributions $\rho_A(f|N)$ shift to larger forces.
For large $\mu$ rebinding becomes irrelevant and the system appears to be irreversible.
In this limit it is of course not possible to gather information about the open state $B$.

\begin{figure}[h!]
\center
\includegraphics[width=9.0cm]{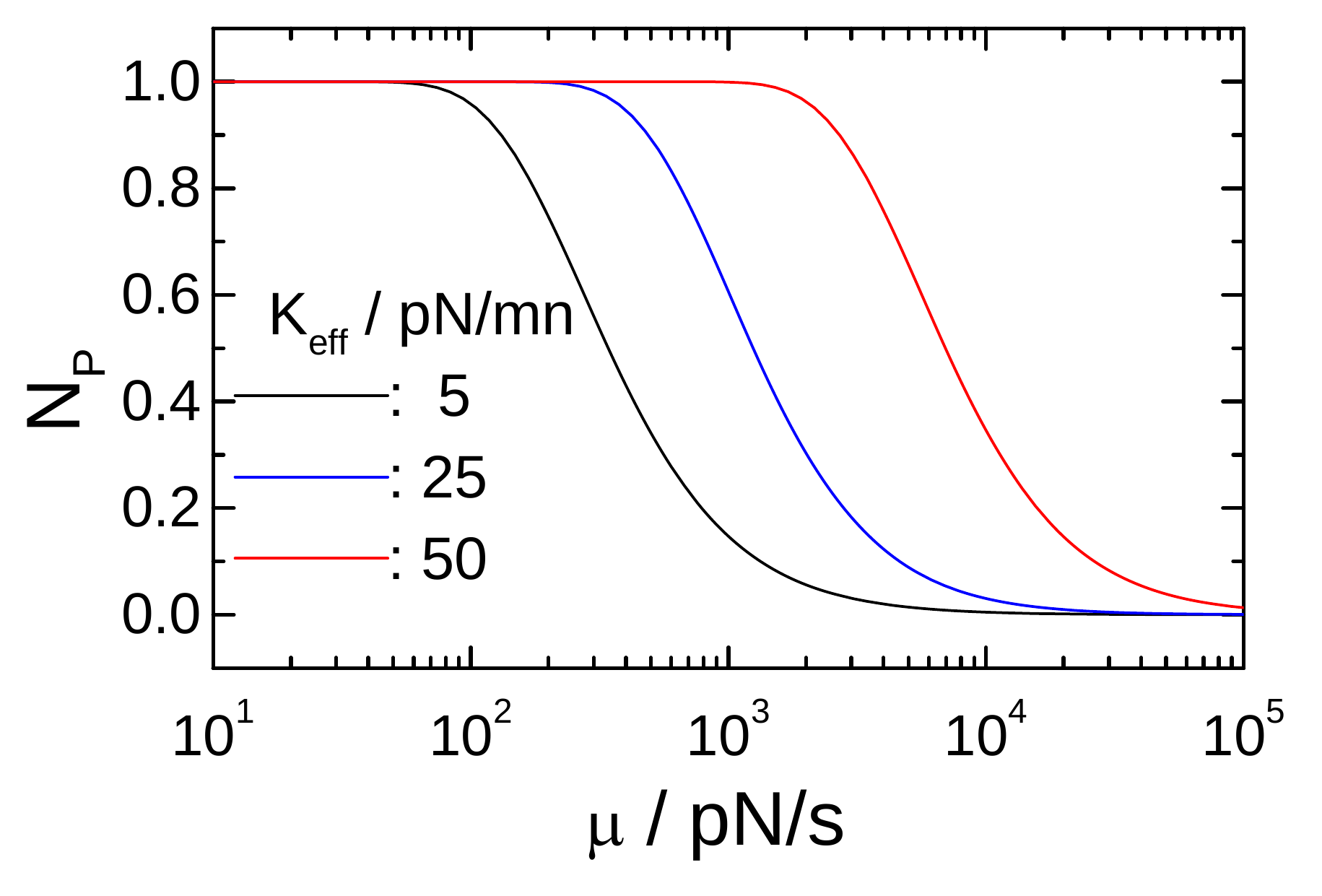}
\vspace{-0.25cm}
\caption{Fraction of trajectories showing at least one rebinding event, $N_P$, as a function of loading rate for various values for $\Keff=5,\, 25,\, 50$ pN/nm.
The remaining parameter are the same as in Fig.\ref{Fig.one}.}
\label{Fig.two}
\end{figure}

In order to quantify the impact of reversibility on the FE curves it is useful to have some measure of the number of back and forth transitions. 
For example, one can compare the first and the last transition force of a given FE-curve as has been done by Evans and Ritchie\cite{Evans:1997}.
A more quantitative measure can be obtained when considering the fraction of trajectories that show at least one rebinding transition, $N_P$.
This quantity has been introduced by Manosas and Ritort\cite{Manosas:2005} and can be cast in the form:
\be\label{NP.def}
N_P=\int_0^{f_e}\!\!df{\partial G_A(f,f_0)\over\partial f}\int_{f}^{f_e}\!\!df'{\partial G_B(f',f)\over\partial f'}
=1-P_A(0|f_e)-P_B(1|f_e)
\ee
where it is assumed that the trajectories start at $f=0$ and that the final force $f_e$ is large enough to find the system in state $B$ with unit probability.
If monitored as a function of the final force, one can deduce $\rho_B(f|1)$ from $N_P(f)$ since 
$\partial N_P(f)/\partial f=-\rho_B(f|1)$.
In Fig.\ref{Fig.two}, $N_P$ is shown for different $\Keff$.
It is found that rebinding at a given loading rate is more important for larger $\Keff$.
This is because $\a_B$ decreases with increasing $\Keff$ and therefore the force dependence of $k_B(f)$ is weaker.
This results in a larger force regime with $k_B(f)/k_A(f)$ and therefore more back and forth transitions.

As has been mentioned already in the previous section, for small loading rates (large $N_P$) the fluctuations complicate a straightforward analysis of the FE-curves and it is meaningful to consider averaged FE curves (also termed 'dynamic strength\cite{Seifert:2002}). 
From these curves, the average rupture forces can be extracted.
In order to relate these average forces to the populations $p_X(f)$, we define a stationary rupture force distribution in the way already indicated above. 
According to eg.(\ref{pX.Sum.PXN}), the force-dependent populations can be obtained from summing up all $P_X(N|f)$. 
Using eq.(\ref{RhoXN.PXN.TSM}), we can thus define
\be\label{rhoX.eq}
\rho_X^{(st.)}(f)=\sum_{n=0}^\infty\rho_X(f|n)=k_X(f)p_X(f)
\ee
Note that this definition differs from those given earlier\cite{G67, Chen:2005}.
In ref.\cite{G67}, we used $(-d p_X(f)/df)$ as a definition for a rupture force distribution.
Without showing results here, I mention that there are only marginal differences to the behavior of 
$\rho_X^{(st.)}(f)$ and that none of the conclusions conducted in ref.\cite{G67} are affected by this change in definition.
It should be mentioned that the relation of $\rho_X^{(st.)}(f)$ to the populations $p_X(f)$ holds only for the TSM due to eq.(\ref{RhoXN.PXN.TSM}). 
If deviations from Markovian behavior are considered, the relation will usually be more involved.

Eq.(\ref{rhoX.eq}) allows to compare the mean rupture forces 
\be\label{Feq.F0}
\lg F\rg_A^{(st.)}=\int\!df\cdot f\cdot\rho_A^{(st.)}(f)
\quad\mbox{and}\quad
\lg F\rg_A^{(0)}=\int\!df\cdot f\cdot\rho_A(f|0)
\ee
While $\lg F\rg_A^{(st.)}$ depends on the populations of both wells and therefore on the statistics of the reversible transitions $\lg F\rg_A^{(0)}$ is independent of any back-transition from well $B$ once the system has left well $A$.
\begin{figure}[h!]
\center
\includegraphics[width=9.0cm]{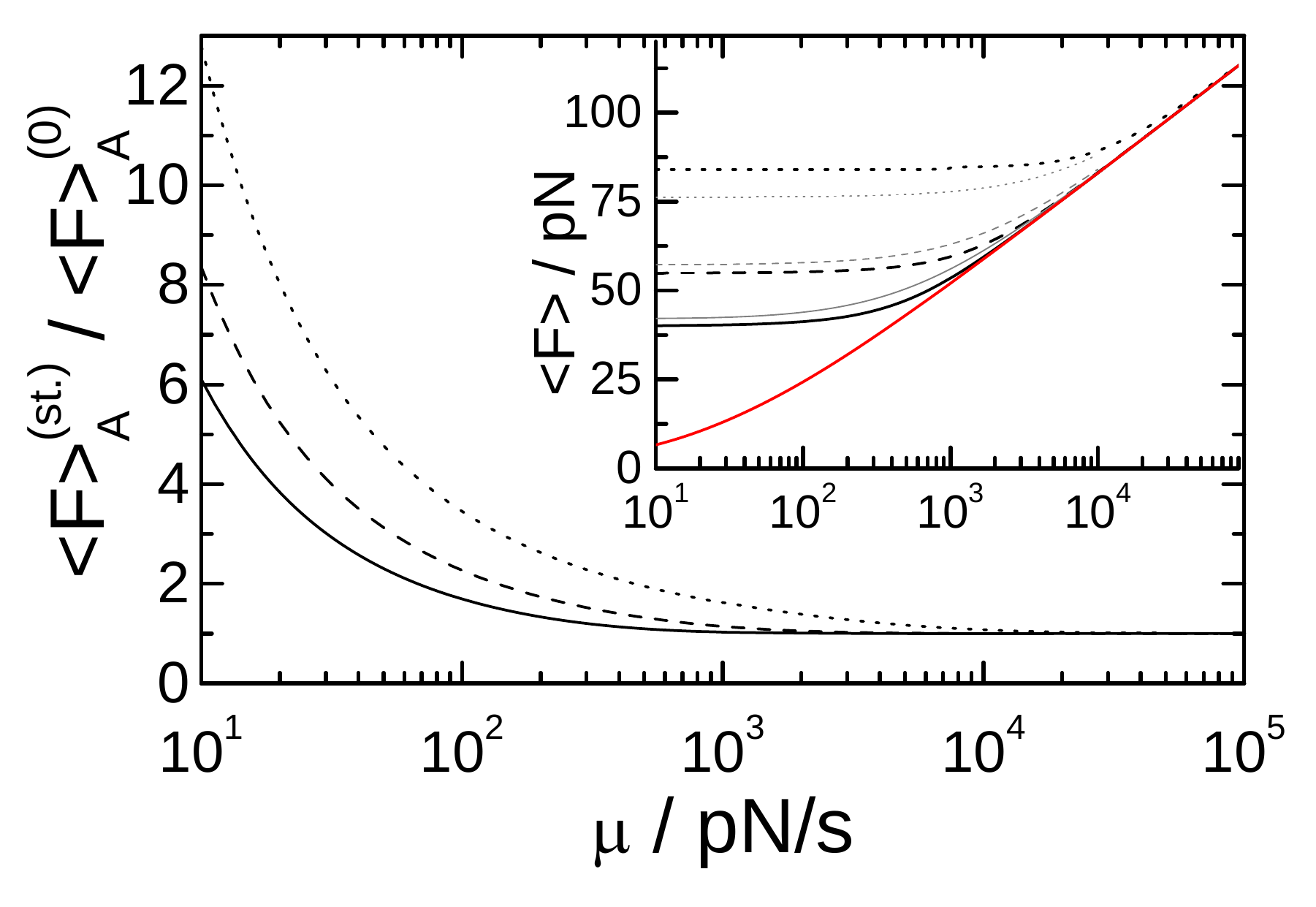}
\vspace{-0.25cm}
\caption{Ratio of $\lg F\rg_A^{(st.)}$ and $\lg F\rg_A^{(0)}$ as a function of loading rate for different values of $\Keff$; full line: $\Keff=5$ pN/nm, dashed line: $25$ pN/nm, dotted line: $50$ pN/nm.
The inset shows $\lg F\rg_A^{(st.)}$ (black lines) and $\lg F\rg_A^{(0)}$ (red line).
The light thin lines are obtained using the phenomenological definition $(-d p_X(f)/df)$ given in ref.\cite{G67} for the distribution.
The remaining parameter are the same as in Fig.\ref{Fig.one}.}
\label{Fig.three}
\end{figure}
The ratio of these quantities is plotted in Fig.\ref{Fig.three}.
The inset shows the individual rupture forces $\lg F\rg_A^{(st.)}$. 
It is apparent that rebinding is very important for small loading rates and that for larger $\Keff$ (or smaller 
$\a_B$) higher loading rates are required in order to reach the irreversible limit.
This observation holds generally: Soft springs and large loading rates give rise to a vanishing rebinding probability.
For stiff springs and small loading rates reversible rebinding can be observed.
These observations are in accord to earlier findings\cite{G67, Tshiprut:2008, Maitra:2010}.

The inset in Fig.\ref{Fig.three} shows the mean forces and it is apparent that $\lg F\rg_A^{(st.)}$ approaches a finite strongly $\Keff$-dependent equilibrium value for small loading rates, as has been observed earlier\cite{G67, Friddle:2008}.
Note that in contrast to $\lg F\rg_A^{(st.)}$ the quantity $\lg F\rg_A^{(0)}$ is independent of $\Keff$
due to the chosen parameters with $\KmA\gg\Keff$.
In general, also $\lg F\rg_A^{(0)}$ might depend on the stiffness of the pulling device.
When comparing the average forces to $N_P$ (Fig.\ref{Fig.two}) it is evident that $N_P$ approaches zero for a loading rate that is similar to the one where $\lg F\rg_A^{(st.)}$ and $\lg F\rg_A^{(0)}$ approach each other.

In order to extract more detailed informations about the kinetics of state $B$ from individual FE-curves, it is meaningful to consider the probability $\rho_B(f|1)$, i.e. the probability to find the first rebinding event for force $f$.
This quantity can be computed from eqns.(\ref{PA0.PB1}, \ref{RhoXN.PXN.TSM}) and is explicitly given by: 
\be\label{rhoB1f}
\rho_B(f|1)=\mu^{-1}k_B(f)P_B(1|f)
=\mu^{-2}k_B(f)e^{Z_A(0)+Z_B(f)}\int_0^f\!df'k_A(f')e^{-[Z_A(f')+Z_B(f')]}
\ee
where $Z_X(f)=k_X(f)/(\a_X\mu)$.
\begin{figure}[h!]
\center
\includegraphics[width=9.0cm]{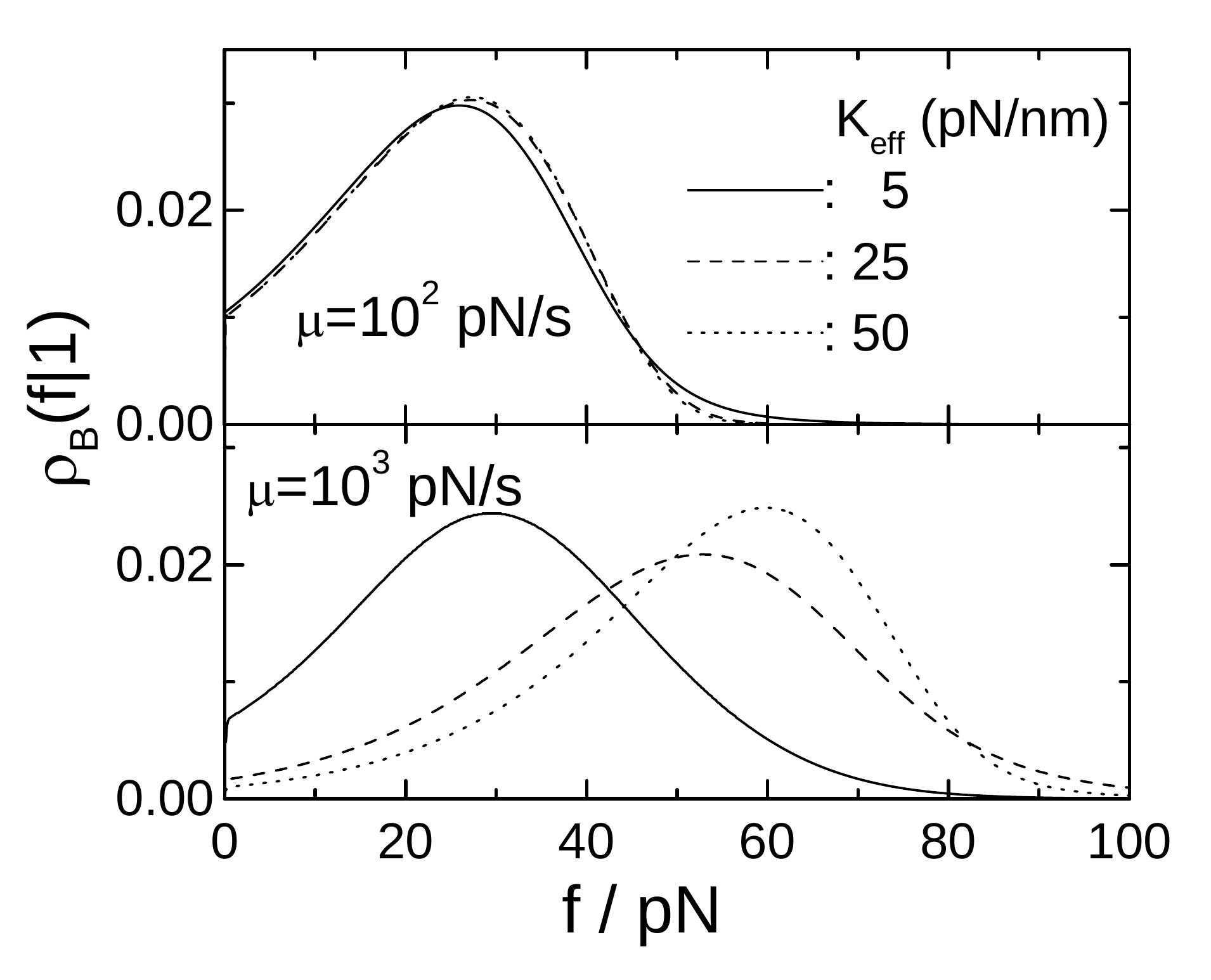}
\vspace{-0.25cm}
\caption{Normalized distributions $\rho_B(f|1)$ versus force for various values for $\Keff=5,\, 25,\, 50$ pN/nm corresponding to
$\a_B/\b=0.61$, $0.37$ and $0.2$, respectively. 
The remaining parameter are the same as in Fig.\ref{Fig.one}.
}
\label{Fig.four}
\end{figure}
In Fig.\ref{Fig.four}, the normalized distribution $\rho_B(f|1)$ is shown for two values of the loading rate and different 
$\Keff$ corresponding to different $\a_B$.
Note that for larger $\a_B$, the kinetic rate $k_B(f)$ shows a stonger decrease with force, cf. eq(\ref{KX.Bell}).
It is evident from Fig.\ref{Fig.four} that for larger loading rates a separation of the distributions for small and larger 
$\Keff$ is observable.
For small values of $\mu$ the distributions are very similar almost independent of $\Keff$ and this changes for larger loading rates.
The distribution for small $\Keff$ (large $\a_B$) does only slightly move to higher forces with increasing $\mu$ in contrast to the other examples for larger $\Keff$.
Thus, a variation of $\Keff$ allows to determine kinetic parameters when fitting $\rho_B(f|1)$ to eq.(\ref{rhoB1f}).
Note that this equation has two fitting parameters, if $k_A(0)$ and $\a_A$ are determined from the rupture force distribution
$\rho_A(f|0)$.

\begin{figure}[h!]
\center
\includegraphics[width=7.15cm]{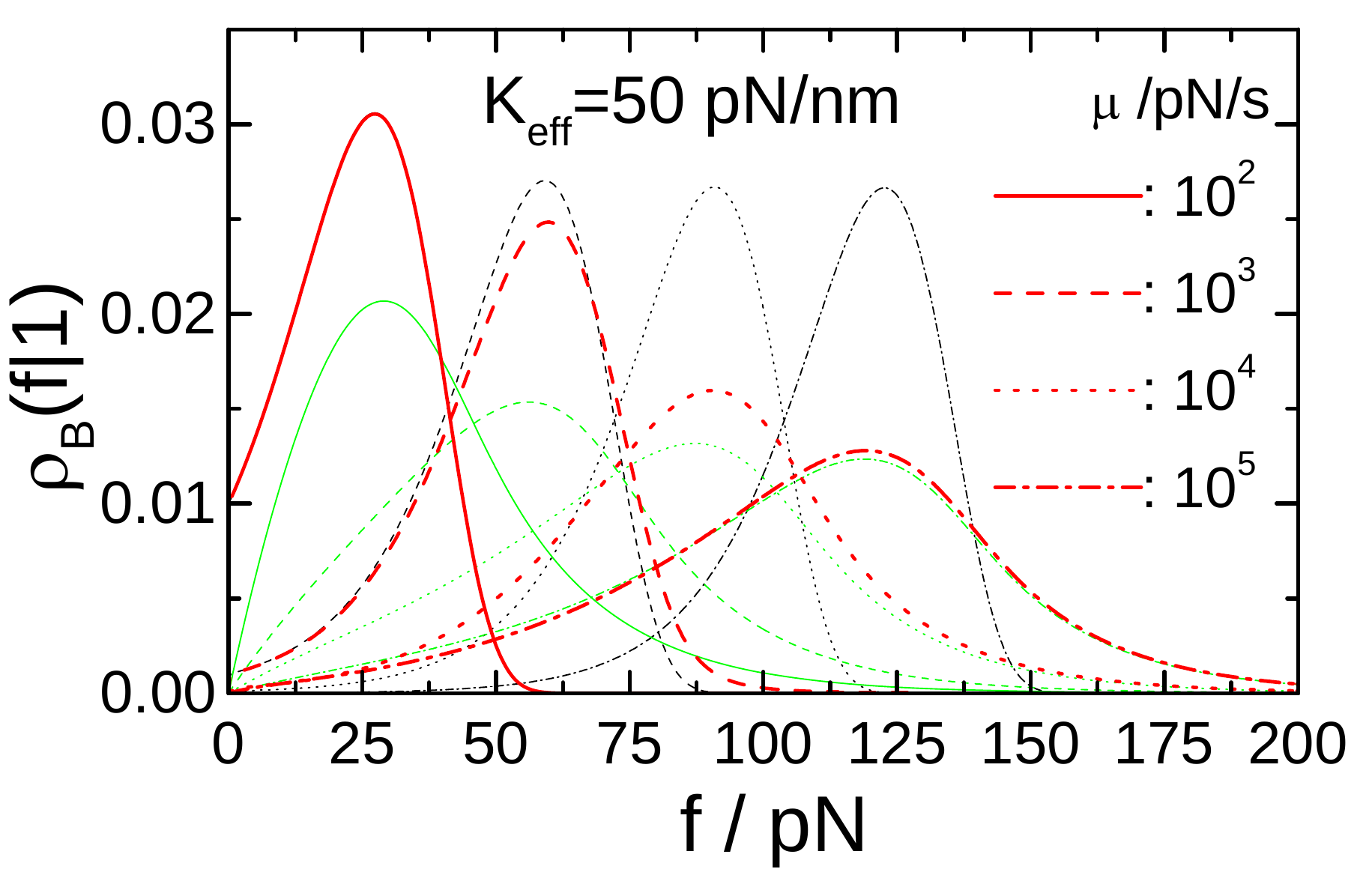}
\includegraphics[width=7.0cm]{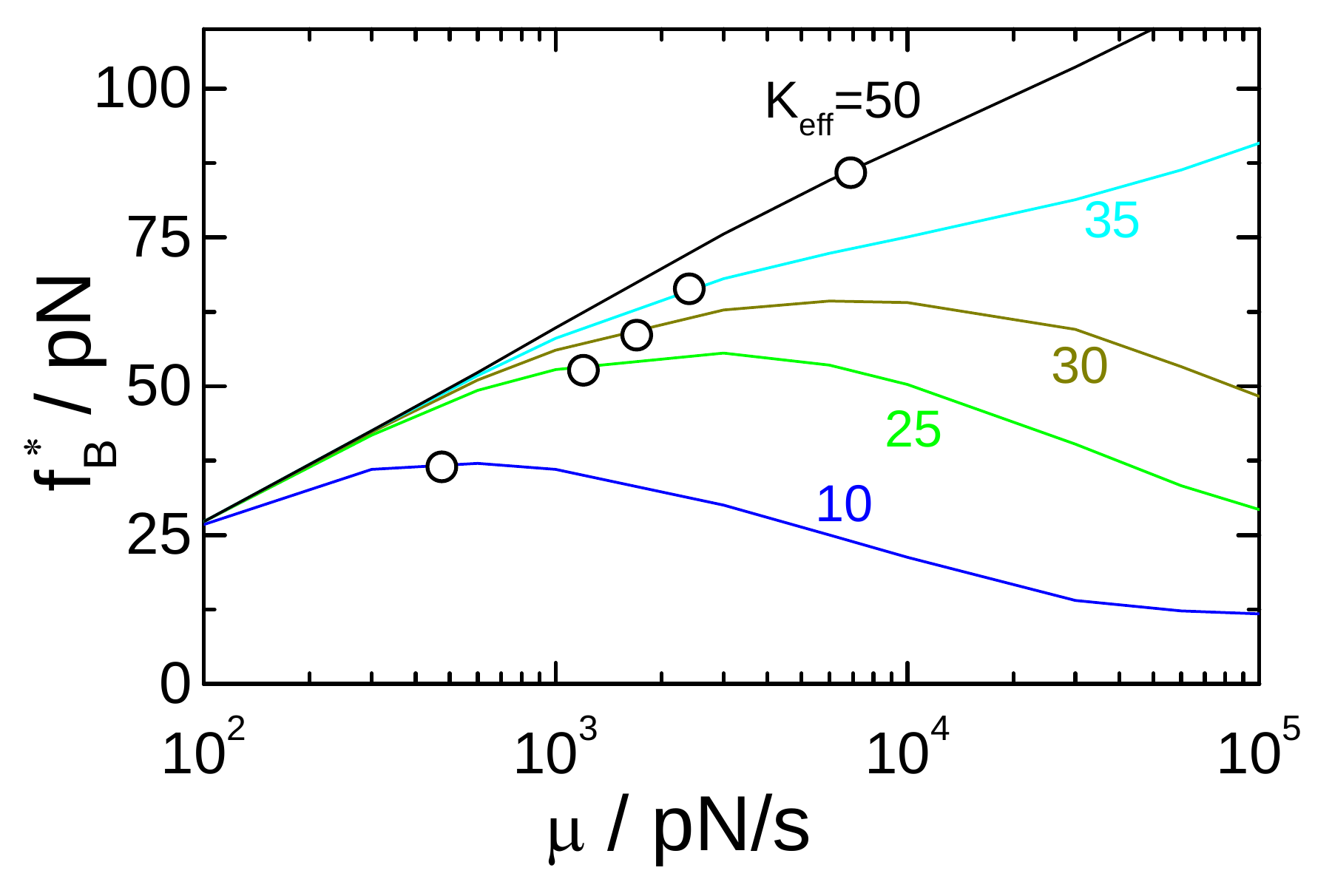}
\vspace{-0.25cm}
\caption{a) (left) Normalized distributions $\rho_B(f|1)$ versus force for different loading rates with $\mu$ (pN/s): $10^2$ full lines, $10^3$ dashed, $10^4$ dotted and $10^5$ dash-dotted.
The thick red lines represent the exact results, upper thin black lines the limit of small $\mu$ and lower thin green lines the limit of large $\mu$.
b) (right) Values of the force at the maximum of $\rho_B(f|1)$, $f^*_B$, as a function of $\mu$ for various 
$\Keff$ as indicated.
The open circles indicate the value of $\mu$ below which rebinding is expected to be important, defined by $N_P=1/2$.
The remaining parameter are the same as in Fig.\ref{Fig.one}.}
\label{Fig.five}
\end{figure}

The fact that for small loading rates $\rho_B(f|1)$ appears independent of $\Keff$ can be understood if one considers the limiting behavior of $\rho_B(f|1)$ for small and large values of $\mu$, which is obtained starting from eq.(\ref{rhoB1f}).
For small $\mu$, one uses the fact that for small forces one has $k_A(f)\ll k_B(f)$ and for large forces the opposite behavior holds.
Using this in the calculation of $P_B(1|f)$, one finds:
\begin{align}\label{PB1.mu.small.large}
\rho_B(f|1)&\simeq \mu^{-1}k_A(f)G_A(f,f_0)\equiv\rho_A(f|0)
\quad;\quad \mu\to0\\
\rho_B(f|1)&\simeq \mu^{-1}k_B(f)\left[1-G_A(f,f_0)\right]
\quad;\quad \mu\to\infty
\nonumber
\end{align}
It is evident that for small $\mu$ the behavior is completely independent of the properties of state $B$ and only for larger $\mu$ information about the $k_B(f)$ can be obtained.
This behavior is exemplified in Fig.\ref{Fig.five}a, where the $\mu$-dependence of $\rho_B(f|1)$ is shown including the limiting behavior.
For small $\mu$ one has the limiting behavior that is independent of $k_B(f)$.
For larger $\mu$ the behavior strongly depends on $\a_B$ and $\Keff$ and for large $\mu$ the corresponding limit is recovered.
The value of the force at the maxima of the distributions, $f^*_B$, is shown in Fig.\ref{Fig.five}b.
There, the value at which $N_P=1/2$ is indicated as the open circle.
This value might be taken as an indication of the maximum loading rate for which a finite detectable number of rebinding transitions can be expected to occur in an FE-curve.
Thus, as already discussed above in the context of $N_P$ there is a strongly $\Keff$-dependent $\mu$-range in which it is possible to extract meaningful information about the rebinding kinetics from the distributions $\rho_B(f|1)$.

\begin{figure}[h!]
\center
\includegraphics[width=7.25cm]{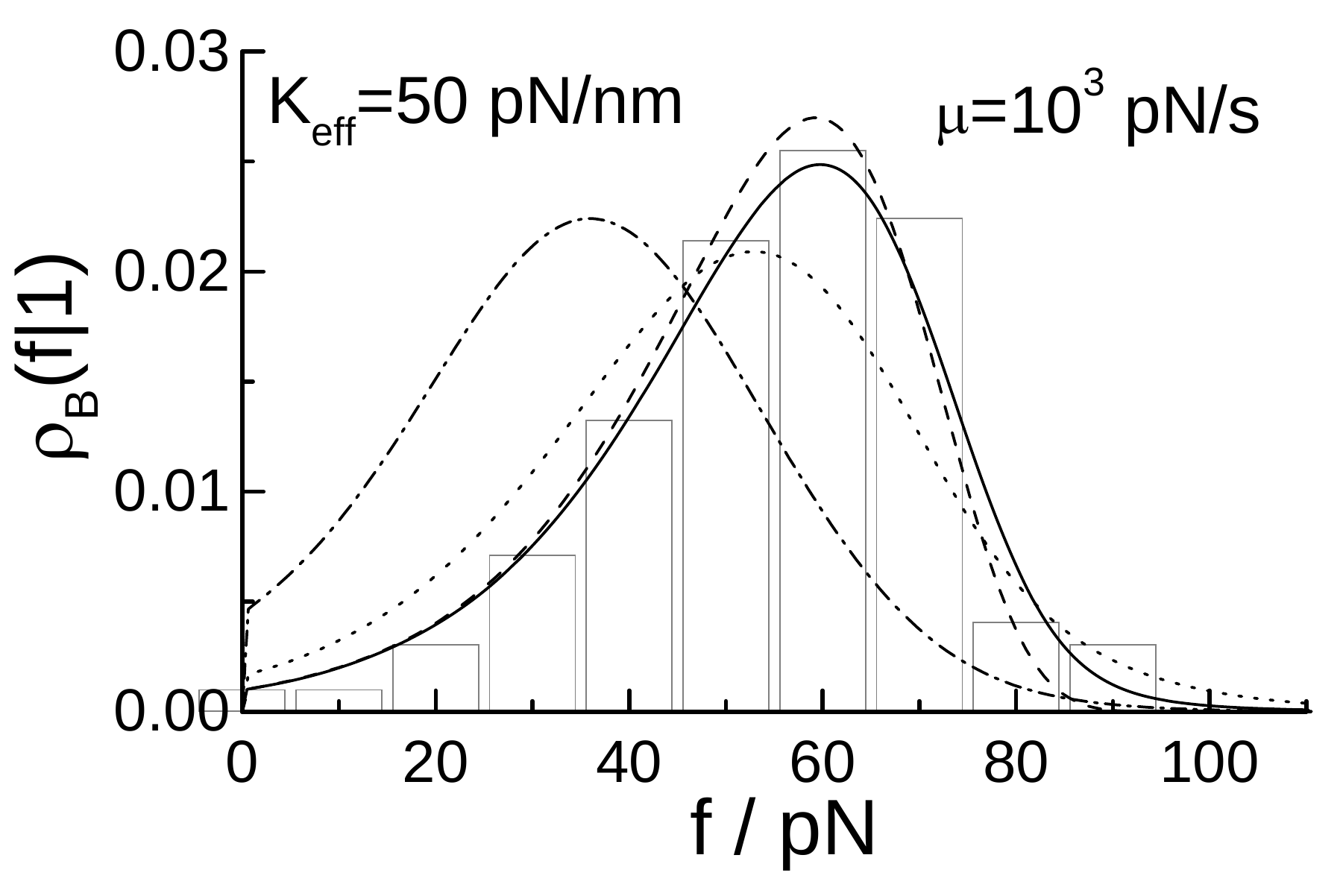}
\includegraphics[width=7.15cm]{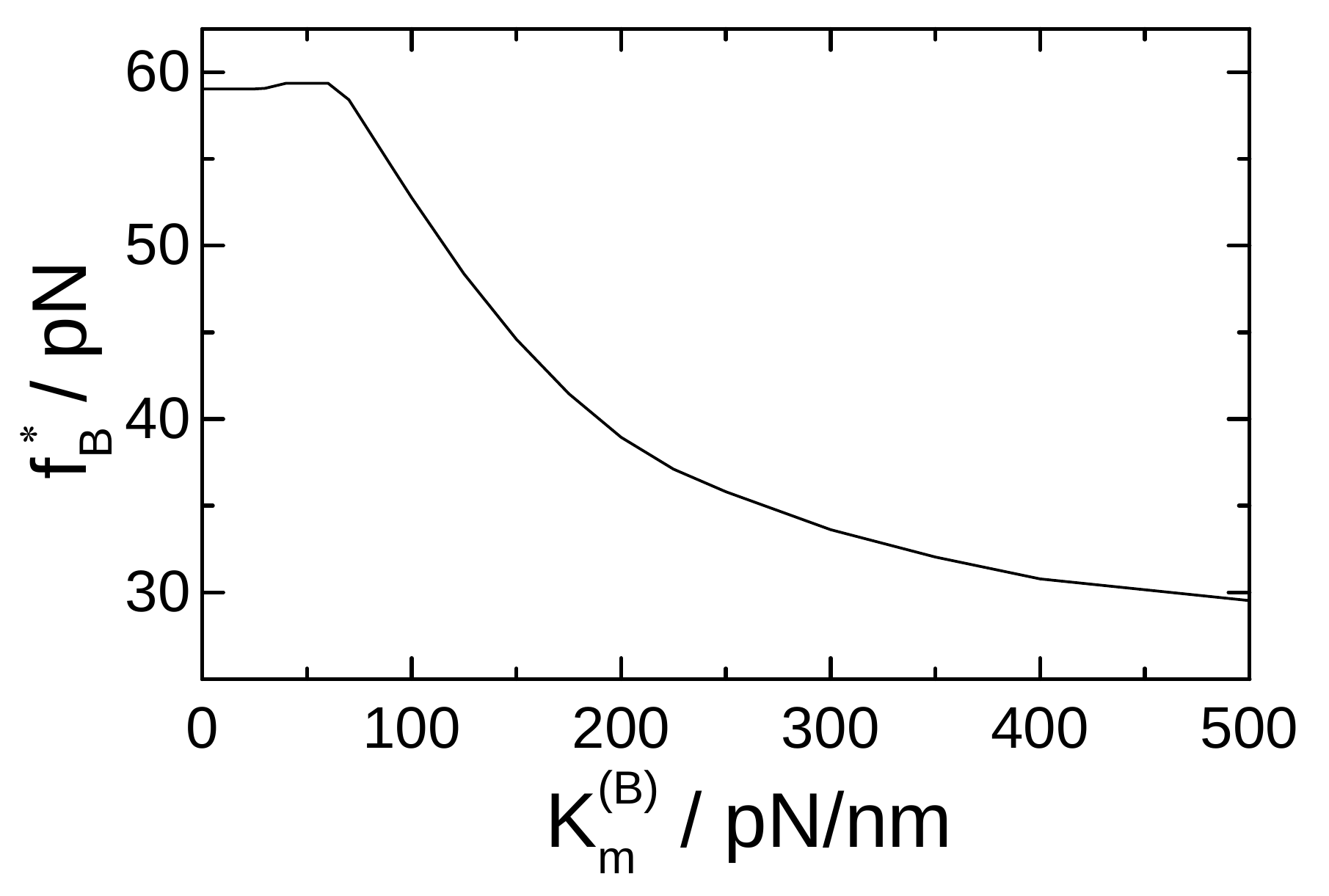}
\vspace{-0.25cm}
\caption{a) (left panel) Normalized distributions $\rho_B(f|1)$ versus force for different molecular force constants $\KmB$ (pN/nm): $50$: full line, 
$25$: dashed, $100$: dotted and $250$: dash-dotted. The thin bars are results of KMC simulations for $\KmB=50$ (pN/nm) and 100 observed transitions for comparison.
b) (right panel) $f^*_B$ as a function of $\KmB$.
The remaining parameter are the same as in Fig.\ref{Fig.one}.}
\label{Fig.six}
\end{figure}

In order to exemplify the dependence of $\rho_B(f|1)$ on the parameters, in Fig.\ref{Fig.six}a this quantity is shown for different value of $\KmB$.
Here, $\Keff=50$ pN/nm and $\mu=10^3$ pN/s, which are parameters for which a significant number of rebinding transitions are to be expected, $N_P\simeq1$, cf. Fig.\ref{Fig.two}.
The thin bars indicate the distribution $\rho_B(f|1)$ as obtained from KMC simulations for the parameters used in Fig.\ref{Fig.one} for a data set with 100 FE-curves that exhibit a first rebinding transition, which can be expected to be observable experimentally.
It is apparent from Fig.\ref{Fig.six}a that the dependence on $\KmB$ (and accordingly on $\a_B$) should be observable.
In Fig.\ref{Fig.six}b $f^*_B$ is plotted as a function of $\KmB$. 
It shows that this value changes by a factor of two from the limit $\KmB\ll\Keff$ to the opposite, $\KmB\gg\Keff$. 

As we have already discussed in some detail in ref.\cite{G67}, it is also recommendable to analyze the so-called dynamic strength, i.e. averaged FE-curves, eq.(\ref{dyn.strength}).
While the loading rate dependence of $\lg F(f)\rg$ has been discussed in ref.\cite{G67}, it should be mentioned that averaged FE-curves also might yield valuable information for the determination of the kinetic rates and of $\KmB$.
This is, as discussed above, because the slope of the FE-curves in the open $B$ state is given by $\xi_B\Keff$ and thus depends on $\KmB$.
Additionally, the populations $p_X(f)$ enter the expression for the Gaussian approximation and this fact allows to extract informations regarding the kinetic rates.

The following procedure can therefore be used to obtain valuable information about the rebinding kinetics of reversibly bound systems from the FE curves.
Fitting the distribution of first rebinding forces, $\rho_B(f|1)$ to eq.(\ref{rhoB1f}) allows to extract the parameters 
$\a_B$ and $k_B(0)$ and from $\a_B$ one has access to the elastic constant $\KmB$.
 A consistency check can be performed if the obtained values are used to compute the dynamics strength, in particular if the FE curves are recorded for different stiffnesses of the pulling device, $\Keff$.
\section*{IV. Conclusions}
The kinetics of systems exhibiting reversible rebinding after bond rupture can be investigated using force spectroscopy.
In the standard application of force ramp spectroscopy the molecular bond is opened and the rupture force distribution gives information about the kinetics of the transition from the closed to the open state.
If no fluctuations are present, it is usually not possible to extract information about the open state and about the rebinding kinetics. 
This situation changes if it is possible to perform force clamp spectroscopy.
In this case, the waiting time distributions can be analyzed and this yields the desired information.
Similar information can be conducted if in the force ramp case one can observe rebinding in the relax mode, i.e. via decreasing the distance of the pulling device and the substrate and thus inverting the pulling direction.
This latter possibility often is hampered by the presence of soft linkers which prevent the molecular complex to reorganize such that rebinding can occur.
In this case, one often can only observe some transient rebinding transitions in the pull mode.
These rebinding transitions appear as fluctuations in the FE-curves and occur in the force range where the equilibrium constant defined by the kinetic rates is on the order of unity.

In the present paper, I developed the theory necessary for describing force ramp spectroscopy in case of relevant rebinding. 
This is a straightforward generalization of the standard theory of fluctuations in two-state systems for the case of time-independent kinetic rates. 
Using the various distributions that can be defined, I have shown how a meaningful definition of a stationary rupture force distribution can be given also in case of finite rebinding probability if deviations from Markovian behavior are negligible.
This stationary rupture force distribution is proportional to the population of the bound state but it differs from the phenomenological definition that has been used earlier\cite{G67}. 
The differences between the resulting distributions and mean forces, however, are only of a quantitative manner.
An extension of the treatment to the case of systems exhibiting dynamic disorder will be undertaken in a future study.

I have shown that the analysis of the distribution of the first rebinding transitions extracted from FE-curves conducted in the pull mode can provide valuable information about the energy landscape of the open state and about the kinetics of the rebinding transition.
In the actual model calculations, I used the phenomenological Bell model in a slightly generalized form allowing to include a dependence of the kinetic rates on the stiffness of the pulling device, $\Keff$.
I have shown that the variation of $\Keff$ can be extremely useful in the determination of the parameters describing the energy landscape of the system.
In the present paper I have disregarded a dependence on $\Keff$ of the unbinding rate, but this can easily be included additionally.
Also the analysis of averaged FE-curves for different $\Keff$ is expected to allow the extraction of parameters determining the kinetic rates for the unbinding and the rebinding transitions.

In summary, I have developed the stochastic theory for the treatment of two-state systems with time-dependent rates and I have applied this theory to the case of force ramp spectroscopy.
It has been found that even if it is not possible to detect the rebinding in a relax mode experiment, informations about the rebinding rate and the energy landscape of the open state can be obtained from the fluctuations in the FE-curves.
The stiffness of the pulling device has been shown to be a very important additional experimental parameter in the determination of the properties of reversibly bound systems.
\section*{Acknowledgement}
I thank Thomas Schlesier, Stefan Jaschonek, Burkhard Geil and Andreas Janshoff for fruitful discussions.
Financial support by the Deutsche Forschungsgemeinschaft via the SFB 625 (A8) is acknowledged.
%
\begin{appendix}
\section*{Appendix A: Generating functionals for $P_X(n|t)$ and $\rho_X(t|n)$}
\setcounter{equation}{0}
\renewcommand{\theequation}{A.\arabic{equation}}
One way to relate the two quantities $P_X(n|t)$ and $\rho_X(t|n)$ consists in considering the corresponding generating 
functional\cite{GS:2006,vkamp}:
\be\label{F.Ftilde.def}
F_X(z,t)=\sum_{n=0}^\infty z^nP_X(n|t)
\quad\mbox{and}\quad 
\tilde F_X(z,t)=\sum_{n=0}^\infty z^n\rho_X(t|n)
\ee
However, since both quantities depend on the matrix $\P_X(n|\t,t_0)$, it is more convenient in the present context to consider:
\be\label{Gzt.PN}
\mathbf\Gamma_X(z|t,t_0)=\sum_{n=0}^\infty z^n\P_X(n|t,t_0)
\ee
Using the master equation, eq.(\ref{ME.AB}), and eq.(\ref{PXN.def}), one can show that 
$\mathbf\Gamma_X(z|t,t_0)$ obeys:
\be\label{Gz.eom}
{d\over dt}\mathbf\Gamma_X(z|t,t_0)=\left[\W'(t)+z\V(t)\right]\mathbf\Gamma_X(z|t,t_0)
\ee
With this matrix at hand, we have for the generating functionals $F(z,t)$ and $\tilde F(z,t)$:
\be\label{F.Ftilde.res}
F_X(z,t)=\ProjP_X\left\{\mathbf\Gamma_X(z|t,t_0)\P(t_0)\right\}
\quad\mbox{and}\quad
\tilde F_X(z,t)=\ProjP_X\left\{\V(t)\mathbf\Gamma_X(z|t,t_0)\P(t_0)\right\}
\ee
Note that from eq.(\ref{Gz.eom}) it follows that $F_X(z=1,t)=p_X(t)$.
While the $P_X(n|t)$ are obtained as derivatives of $F_X(z,t)$ with respect to $z$ in the standard way, it is not possible to use 
$\tilde F(z,t)$ in the same way for the computation of $\rho_X(t|n)$.
As has been pointed out in a similar discussion in ref.\cite{GS:2006} this is because $\tilde F(z,t)$ is not a generating functional in the usual sense.
However, using the decomposition of the transition matrix given in eq.(\ref{VW.events}), one finds the following relation between the two functionals:
\Be\label{F.relations}
\dot F_X(z,t)=
&&\hspace{-0.6cm}
-\tilde F_X(z,t)+z\tilde F_Y(z,t)
\nonumber\\
(1-z^2)\tilde F_X(z,t)=
&&\hspace{-0.6cm}
-\dot F_X(z,t)-z\dot F_Y(z,t)
\Ee
Inserting the definition of $F(z,t)$ and $\tilde F(z,t)$, eq.(\ref{F.Ftilde.def}), into these expressions one obtains 
eqns.(\ref{PNX.aus.RhoNX}), (\ref{RhoNX.aus.PNX})  in the text.

\section*{Appendix B: Transition rates in a local harmonic approximation}
\setcounter{equation}{0}
\renewcommand{\theequation}{B.\arabic{equation}}
Here, I give the expressions for the Kramers rates in a general double well potential\cite{G67}.
Assuming that the reaction coordinate $q$ shows minima located at $q_A$ and $q_B$ and a transition state at $q_T$ that can be approximated locally by parabola with curvatures curvatures $\KmA$, $\KmB$ for the minima and $\KmT$ for the maximum, one finds:
\be\label{koff.Kramers}
k_A(f)=k_A(0)e^{\a_Af\left[1-{1\over 2}f/f_c\right]}
\quad\mbox{and}\quad
k_B(f)=k_B(0)e^{-\a_Bf\left[1+{1\over 2}f/f_r\right]}
\ee
With the definitions:
\be\label{XiX.def}
\xi_X=\left[1+\Keff/\KmX\right]^{-1}\,\,(X=A,\,B)\,\,
\quad\mbox{and}\quad
\xi_T=\left[1-\Keff/\KmT\right]^{-1}
\ee
one has
\be\label{alfa.LH}
\a_A=\b\left(\xi_T q_T-\xi_Aq_A\right)
\quad;\quad
\a_B=\b\left(\xi_Bq_B-\xi_Tq_T\right)
\ee
Furthermore, the scale of the forces are set by
\be\label{fc.fr}
f_c={\KmT\KmA\over\xi_A\KmT+\xi_T\KmA}\left({\a_A\over\b}\right)
\quad\mbox{and}\quad
f_r={\KmT\KmB\over\xi_B\KmT+\xi_T\KmB}\left({\a_B\over\b}\right)
\ee
In the text, it is assumed that forces are small, $f/f_c\ll1$ and $f/f_r\ll1$.
Additionally, the barrier is approximated as a sharp barrier, $\KmT\gg\Keff$ and $\xi_T\simeq1$, and the minimum of the 'closed' configuration is deep and has a curvature that is much larger than the stiffness of the pulling device, $\KmA\gg\Keff$ or $\xi_A\simeq1$.
In this case, the rates can be written as in eq.(\ref{KX.Bell}) with
\be\label{alfaX.small}
\a_A=\b (q_T-q_A)
\quad\mbox{and}\quad
\a_B=\b\left(\xi_Bq_B-q_T\right)
\ee
and the critical force reduces to the value of vanishing barrier, $f_c\simeq \KmA(q_T-q_A)$.
\\

The expression for the mean force as a function of extention, the so-called dynamic strength, reads in a Gaussian aproximation, 
cf. \cite{G67}:
\be\label{dyn.strength}
\lg F(x)\rg=\xi_A\Keff\left(x-q_A\right)p_A(f)+\xi_B\Keff\left(x-q_B\right)p_B(f)
\ee
\end{appendix}
\newpage
\end{document}